\documentclass[a4paper,11pt,floatfix,showpacs,tightenlines,showkeys,superscriptaddress,amsmath,amssymb,nofootinbib]{article}
\pdfoutput=1
\usepackage{jcappub}
\usepackage{amssymb,amsbsy,epsfig,color,graphicx}
\usepackage{color}
\usepackage{[longtable}
\usepackage{array}
\usepackage{dcolumn}   
\usepackage{cellspace}
\usepackage{mathtools}
\usepackage{amstext}
\usepackage{amssymb}
\usepackage{stmaryrd}
\usepackage{stackrel}
\usepackage{graphicx}
\usepackage{esint}
\usepackage[utf8]{inputenc}
\usepackage{blindtext}
\usepackage{float}
\restylefloat{table}
\usepackage{booktabs}
\usepackage{enumitem} 

\usepackage{etoolbox} 
\usepackage{lipsum} 
\usepackage[capitalize]{cleveref}

\usepackage{multirow}
\usepackage[caption=false]{subfig}
\renewcommand\[{\begin{equation}}
\renewcommand\]{\end{equation}}

\newcommand{\ba}{\begin{eqnarray}}
\newcommand{\ea}{\end{eqnarray}}

\makeatletter

\appto{\appendix}{%
	\@ifstar{\def\thesection{\unskip}\def\theequation@prefix{A.}}%
	{\def\thesection{\Alph {section}}}%
}
\makeatother

\begin{document}

\begin{flushleft}
KCL-PH-TH/2020-70 
\end{flushleft}
\begin{flushleft}
CERN-TH-2020-198
\end{flushleft}
	
	\title{Constraints on extended gravity models through gravitational wave emission}

	\author[a,b]{Gaetano Lambiase}
	\author[c,d]{, Mairi Sakellariadou}
	\author[a,b]{, Antonio Stabile}

	\affiliation[a]{Dipartimento di Fisica "E.R. Caianiello", Universit\`a di Salerno, I-84084 Fisciano (SA), Italy}
	\affiliation[b]{INFN - Sezione di Napoli, Gruppo collegato di Salerno, I-84084 Fisciano (SA), Italy}
	\affiliation[c]{Department of Physics, King's College London, University of London, Strand WC2R 2LS, London, United Kingdom}
	\affiliation[d]{Theoretical Physics Department, CERN, Geneva, Switzerland}
	\date{\today}

\abstract{Using recent experimental results of detection of gravitational waves from the binary black hole signals by Advanced LIGO and Advanced Virgo, we investigate the propagation of gravitational waves in the context of fourth order gravity nonminimally coupled to a massive scalar field. Gravitational radiation admits extra massive modes of oscillation and we assume that the amplitude of these modes is comparable to that of the massless mode. We derive the propagation equation and effective mass for each degree of freedom and we infer, from the current observational data, constraints on the free parameters of the gravity models we considered. In particular, for $f(R)=R-R^2/R_0 $, the constraint obtained from the speed of gravitational waves is not compatible with the one set by Solar System tests, which implies that amplitude of the massive modes could not be detectable with current experiments on Earth.}

\maketitle


\section{Introduction}

The recent observations of gravitational radiation \cite{Abbott1,Abbott2,Abbott3,Abbott4,Abbott5,LIGOScientific:2018mvr,CitGWTC-1,Abbott:2020niy} from binary black holes by the Advanced LIGO and the Advanced Virgo detectors, offered the ultimate confirmation of Einstein's theory of General Relativity (GR) and a novel opportunity to test all models of extended gravity in a strong field dynamical regime. These alternative models to GR have been widely studied and tested to the regime in which the Newtonian and post-Newtonian contributions describe very well the dynamics. In particular, they have  been  constrained by Earth tests (e.g., gravitoelectromagnetism and torsion balance experiments), Solar System tests and by astrophysical data on the variation of the orbital period of binary systems. However, direct detection of gravitational waves from coalescing binary black holes, in an extreme gravitational field regime, offers a new test for extended theories of gravity.

On August 2017, the association of gravitational wave GW170817 \cite{Abbott6} and gamma-ray burst GRB170817A \cite{GoldsteinEtAl} events allowed to verify experimentally, with deviations smaller than a few $10^{-15}$  \cite{Abbott7}, that gravitational waves travel at the speed of light $c$. This measurement has considerable relevance since it provides new insight into fundamental physics and probes the speed of gravitational waves over cosmological distances.

Einstein’s theory of GR has been tested to very high precision in the last decade \cite{-C.-M.}. To date it is the best theory of gravitational interaction, however despite its great success there are still open questions which make the theory incomplete. Several observational data \cite{riess, ast, clo, spe,carrol,sahini} probe that the Universe is spatially flat and is undergoing a period of accelerated expansion. To understand and describe the current astrophysical data two unrevealed ingredients are needed in order to achieve this phenomenology: \textit{dark matter} at galactic and extragalactic scales and\textit{ dark energy} at cosmological scales. The extended gravity models have been considered as a viable theoretical mechanism to explain cosmic acceleration and galactic rotation curves.  In such models, one extends only the geometric sector, without introducing any exotic matter. In such models, gravity shows a different behaviour either above (infrared modification) or below (ultraviolet modification) a certain length scale, whilst the robust results of General Relativity at local and Solar System scales are preserved \cite{Staro80,Starobinsky,BoehmerHarako,BoehmerHarako2,SotiriouFaraoni,FeliceTSUj,Nojiriodin,SilvestriTrodden, DurrerMaartens,SamiNotes,CopelandSami, AmendolaTsujikawa, LambMohantySta, Zubair:2016hhc, Zubair:2016bpi, Zubair:2017oir} (see Refs. \cite{piazza1,piazza2} for a class of models related to Horndeski theory).

Extended theories of gravity have been widely studied in the Newtonian, post-Newtonian and Minkowskian limit   \cite{PRD1,PRD2,FOG_ST,GW_FOG,gravito_mag,anupam,LamSakSta,stabstab,stabstabcap,lamb1,Buoninfante:2019uwo,Buoninfante:2018xiw,
Buoninfante:2018rlq,Buoninfante:2020ctr,Buoninfante:2018xif,Buoninfante:2018lnh,tino,mastro}. In general, one finds modifications of the gravitational potential in Newtonian and post-Newtonian limit, whilst in Minkowskian  limit one obtains massive gravitational wave modes. Many studies, in the weak-field limit, have been conducted in context on various astrophysical systems as the galactic rotation curves, stellar hydrodynamics and gravitational lensing. An interesting application concerns the study of the Casimir effect \cite{lv,Lambiase:2016bjy,Lambiase:2016bjy4, Buchbinder, Bagrov, OdintsovPEDA, Elizalde}.

Corrections to the gravitational Lagrangian were already considered by several authors \cite{Staro80, Starobinsky, BoehmerHarako, BoehmerHarako2, SotiriouFaraoni,  FeliceTSUj, Nojiriodin, SilvestriTrodden, DurrerMaartens, SamiNotes, CopelandSami, AmendolaTsujikawa, LambMohantySta}. From a conceptual viewpoint, there is no reason \emph{a priori} to restrict the gravitational Lagrangian to a linear function of the Ricci scalar minimally coupled to matter.  In particular, one may consider the generalization of $f(R)$ models, where $R$ is the Ricci scalar, through generic functions containing curvature invariants such as the \emph{Ricci squared} ($R_{\alpha\beta}R^{\alpha\beta}$) or the \emph{Riemann squared} ($R_{\alpha\beta\gamma\delta}R^{\alpha\beta\gamma\delta}$), which however are not invariant due to the Gauss-Bonnet invariant $R^2-4R_{\mu\nu}R^{\mu\nu}+R_{\mu\nu\lambda\sigma}
R^{\mu\nu\lambda\sigma}$. Note that the same remark applies to the Weyl invariant $C_{\alpha\beta\gamma\delta}C^{\alpha\beta\gamma\delta}$. Hence, one may add a (massive propagating) scalar field  coupled to geometry; this leads to the {\sl scalar-tensor fourth order gravity}.

Using recent experimental results of detection of gravitational waves from the binary black hole signals by Advanced LIGO and Advanced Virgo,  we will study the propagation of gravitational waves in the context of fourth order gravity nonminimally coupled to a massive scalar field, in the weak field limit (for different models see Refs. \cite{derham,creminelli,gong,GWGB,jain}. In particular, we will impose constraints on the free parameters of extended gravity models from the current observational data.

The outline of this paper is the following. In Section~\ref{STFOG}, we give the action of a {\sl scalar tensor fourth order gravity} and write down the corresponding field equations, which we then solve in the presence of matter within the weak-field approximation. In Section \ref{constraints}, using the recent observations of gravitational radiation from binary black holes by the Advanced LIGO and the Advanced Virgo detectors we put the constraint on the free parameters of the teories. In Section \ref{STFOG_models} ,we analyze some models of extended theory.  We summarize our conclusions  in Section ~\ref{conclusions}.

\section{Scalar Tensor Fourth Order Gravity}\label{STFOG}

Let us consider the action
\begin{eqnarray}\label{FOGaction}
\mathcal{S}\,=\,\int d^{4}x\sqrt{-g}\biggl[f(R,R_{\alpha\beta}R^{\alpha\beta},\phi)+\omega(\phi)\phi_{;\alpha}\phi^{;\alpha}+\mathcal{X}\mathcal{L}_{\rm m}\biggr]~;
\end{eqnarray}
$f$ is an unspecified function of the Ricci scalar $R$, the curvature invariant $R_{\alpha\beta}R^{\alpha\beta}\,\equiv Y$ (where $R_{\mu\nu}$ is the Ricci tensor), and the scalar field $\phi$. The $\mathcal{L}_{\rm m}$ denotes the minimally coupled ordinary matter Lagrangian density, $\omega$ is a generic function of the scalar field $\phi$, and $g$ stands for the determinant of the metric tensor $g_{\mu\nu}$ and $\mathcal{X}\,\equiv\,8\pi G$. Note that unless otherwise specified, we use natural units such that $c=1$ and $\hslash=1$.

In the metric approach, the field equations are obtained by varying the action (\ref{FOGaction}) with respect to $g_{\mu\nu}$:
\begin{eqnarray}\label{fieldequationFOG}
f_RR_{\mu\nu}-\frac{f+\omega(\phi)\phi_{;\alpha}\phi^{;\alpha}}{2}g_{\mu\nu}-f_{R;\mu\nu}+g_{\mu\nu}\Box
f_R+2f_Y{R_\mu}^\alpha
R_{\alpha\nu}&&
\nonumber\\
-2[f_Y{R^\alpha}_{(\mu}]_{;\nu)\alpha}+\Box[f_YR_{\mu\nu}]+[f_YR_{\alpha\beta}]^{;\alpha\beta}g_{\mu\nu}+\omega(\phi)\phi_{;\mu}\phi_{;\nu}&=&
\mathcal{X}\,T_{\mu\nu}~;
\end{eqnarray}
$T_{\mu\nu}\,=\,-\frac{1}{\sqrt{-g}}\frac{\delta(\sqrt{-g}\mathcal{L}_m)}{\delta
g^{\mu\nu}}$ is the the energy-momentum tensor of matter, $f_R\,=\,\frac{\partial f}{\partial R}$, $f_Y\,=\,\frac{\partial f}{\partial Y}$ and $\Box=\nabla_\mu \nabla^\mu = {{}_{;\sigma}}^{;\sigma}$ is the D'Alembertian operator\footnote{The convention for the Ricci tensor is
$R_{\mu\nu}={R^\sigma}_{\mu\sigma\nu}$, while for the Riemann
tensor we define ${R^\alpha}_{\beta\mu\nu}=\Gamma^\alpha_{\beta\nu,\mu}+...$, with the
affinities being the usual Christoffel's symbols of the metric:
$\Gamma^\mu_{\alpha\beta}=\frac{1}{2}g^{\mu\sigma}(g_{\alpha\sigma,\beta}+g_{\beta\sigma,\alpha}
-g_{\alpha\beta,\sigma})$. The adopted signature is $(+---)$.}. The trace of the field equations
(\ref{fieldequationFOG}) reads:
\begin{eqnarray}\label{tracefieldequationFOG}
f_RR+2f_YR_{\alpha\beta}R^{\alpha\beta}-2f+\Box[3
f_R+f_YR]+2[f_YR^{\alpha\beta}]_{;\alpha\beta}-\omega(\phi)\phi_{;\alpha}\phi^{;\alpha}\,=\,\mathcal{X}\,T~;
\end{eqnarray}
$T\,=\,T^{\sigma}_{\,\,\,\,\,\sigma}$ is the trace of
energy-momentum tensor. Varying the action (\ref{FOGaction}) with respect to scalar field $\phi$ we get:
\begin{eqnarray}\label{FE_SF}
2\omega(\phi)\Box\phi+\omega_\phi(\phi)\phi_{;\alpha}\phi^{;\alpha}-f_\phi\,=\,0~,
\end{eqnarray}
where $\omega_\phi(\phi)\,=\,\frac{d\omega(\phi)}{d\phi}$ and $f_\phi\,=\,\frac{\partial f}{\partial\phi}$.

Let us analyse the field equations within the \emph{weak-field approximation} in a
Minkowski background $\eta_{\mu\nu}$:
\begin{equation}\label{PM_me}
g_{\mu\nu}\,\sim\,\eta_{\mu\nu}+h_{\mu\nu}
~~,~~\phi\,\sim\,\phi^{(0)}+\varphi~.\nonumber
\end{equation}
We develop the function $f$ as:
\begin{eqnarray}
\label{LimitFramework}
f(R,R_{\alpha\beta}R^{\alpha\beta},\phi)\,\sim &&f_{R}(0,0,\phi^{(0)})\,R+\frac{f_{RR}(0,0,\phi^{(0)})}{2}\,R^2+\frac{f_{\phi\phi}(0,0,\phi^{(0)})}{2}\,(\phi-\phi^{(0)})^2
\nonumber\\
&&+f_{R\phi}(0,0,\phi^{(0)})R\,(\phi-\phi^{(0)})+f_Y(0,0,\phi^{(0)})R_{\alpha\beta}R^{\alpha\beta}~.
\end{eqnarray}
Note that any other possible contribution to $f$ is negligible~\cite{PRD1,PRD2,FOG_ST}. Using the harmonic gauge condition, $g^{\rho\sigma}\Gamma^\alpha_{\,\,\,\rho\sigma}\,=\,0$, we get $h_{\mu\sigma}^{\,\,\,\,\,\,\,,\sigma}-1/2\,h_{\,,\mu}\,=\,0$, and hence  Eq.~(\ref{fieldequationFOG}) reads \cite{lamb1}:

 \begin{eqnarray}
 \label{PMfieldequationFOG31}
&&(\Box_\eta+{m_Y}^2)\Box_\eta h_{\mu\nu}-\biggl[\frac{{m_R}^2-{m_Y}^2}{3{m_R}^2}\,
\partial^2_{\mu\nu}+\eta_{\mu\nu}\biggl(\frac{{m_Y}^2}{2}+\frac{{m_R}^2+2{m_Y}^2}{6{m_R}^2}\Box_\eta\biggr)\biggr]
\Box_\eta h\nonumber\\
&&\qquad\qquad\qquad\qquad\qquad+2\,{m_Y}^2\,f_{R\phi}(0,0,\phi^{(0)})\,(\partial^2_{\mu\nu}-\eta_{\mu\nu}\Box_\eta)\varphi\,=\,-2\,{m_Y}^2\,\mathcal{X}\,T_{\mu\nu}~,\\\nonumber\\
\label{PMfieldequationFOG32}
&&(\Box_\eta+{m_R}^2)\Box_\eta h+6{m_R}^2\,f_{R\phi}(0,0,\phi^{(0)})\,\Box_\eta\varphi\,=\,2{m_R}^2\,\mathcal{X}\,T~,\\\nonumber\\
\label{PMfieldequationFOG33}
&&(\Box_\eta+{m_\phi}^2)\varphi+\frac{f_{R\phi}(0,0,\phi^{(0)})}{2}\,\Box_\eta h\,=\,0~;
\end{eqnarray}
$\Box_\eta$ is the D'Alambertian operator in flat space, $h\,=\,{h^\sigma}_{\,\sigma}$ and we have set\footnote{One could define a new gravitational constant: $\mathcal{X}\,\rightarrow\,\mathcal{X}\,f_R(0,0,\phi^{(0)})$ and $f_{R\phi}(0,0,\phi^{0})\,\rightarrow\,f_{R\phi}(0,0,\phi^{0})\,f_R(0,0,\phi^{(0)})$.} $f_R(0,0,\phi^{(0)})\,=\,1$ and $\omega(\phi^{(0)})\,=\,1/2$.  The quantities ${m_R}^2, {m_Y}^2, {m_\phi}^2$ are defined\footnote{In our formalism we consider the coefficients of $R^2$ and $\phi ^ 2$ negative. For this reason, inside of Eqs. (\ref{mass_definition}) we have the minus signs. However, the definitions of the masses Eqs. (\ref{mass_definition}) are consistent with the formalism used by K. S. Stelle in Ref. \cite{Stelle}.} as follows:

\begin{eqnarray}\label{mass_definition}
\begin{array}{ll}
{m_R}^2\,\equiv-\frac{f_{R}(0,0,\phi^{(0)})}{3f_{RR}(0,0,\phi^{(0)})+2f_Y(0,0,\phi^{(0)})}~,\\\\
{m_Y}^2\,\equiv\frac{f_{R}(0,0,\phi^{(0)})}{f_Y(0,0,\phi^{(0)})}~,\\\\
{m_\phi}^2\,\equiv-\frac{f_{\phi\phi}(0,0,\phi^{(0)})}{2\omega(\phi^{(0)})}~.
\end{array}
\end{eqnarray}
The field equations (\ref{PMfieldequationFOG31}), (\ref{PMfieldequationFOG32}) and (\ref{PMfieldequationFOG33})  generalise those of Ref.~\cite{GW_FOG}, since in the latter there was no scalar field component. Let us also note that these equations are the weak-field limit of the model discussed in Refs.~\cite{gravito_mag,lamb1}.

To solve Eqs.~(\ref{PMfieldequationFOG31}), (\ref{PMfieldequationFOG32}) and (\ref{PMfieldequationFOG33}) we introduce the auxiliary field $\gamma_{\mu\nu}$, such that:
\begin{eqnarray}\nonumber
(\Box_\eta+{m_Y}^2)\Box_\eta \gamma_{\mu\nu}=\,-2\,{m_Y}^2\,\mathcal{X}\,T_{\mu\nu}~.
\end{eqnarray}
Then the fields $h_{\mu\nu}$ can be written as \cite{lamb1}:
\begin{eqnarray}
\label{Position4}
h_{\mu\nu}= \gamma_{\mu\nu}-\frac{A(\partial)}{m^2_Y}\gamma -2\,{m_R}^2\,f_{R\phi}(0,0,\phi^{(0)})\,B(\partial)\,\varphi~,
\end{eqnarray}
where
\begin{eqnarray}
\nonumber
A(\partial)&=&\biggl[\frac{{m_R}^2-{m_Y}^2}{3}\,\partial^2_{\mu\nu}+\eta_{\mu\nu}\biggl(\frac{{{m_R}^2\,m_Y}^2}{2}+\frac{{m_R}^2+2{m_Y}^2}{6}\Box_\eta\biggr)\biggr] (\Box_\eta+{m_R}^2)^{-1}\,, \nonumber\\
\\
\nonumber
B(\partial)&=&\,\biggl[\Box_\eta^{-1}(\Box_\eta+{m_R}^2)^{-1}\partial^2_{\mu\nu}+\frac{1}{2}\eta_{\mu\nu}(\Box_\eta+{m_R}^2)^{-1}\biggl].
\end{eqnarray}
Using Eq. (\ref{Position4}),  Eqs.~(\ref{PMfieldequationFOG31}), (\ref{PMfieldequationFOG32}) and (\ref{PMfieldequationFOG33}) becomes:
\begin{eqnarray}
\label{PMfieldequationFOG41}
&&(\Box_\eta+{m_Y}^2)\Box_\eta \gamma_{\mu\nu}=\,-2\,{m_Y}^2\,\mathcal{X}\,T_{\mu\nu}~,\\\nonumber\\
\label{PMfieldequationFOG42}
&&(\Box_\eta+{m_Y}^2)\Box_\eta \gamma =\,-2\,{m_Y}^2\,\mathcal{X}\,T~,\\\nonumber\\
\label{PMfieldequationFOG43}
&&\frac{(\Box_\eta+{m_\phi}^2)\varphi}{{m_R}^2f^2_{R\phi}(0,0,\phi^{(0)})} -3\,(\Box_\eta+{m_R}^2)^{-1}\Box_\eta \varphi\,=\,-(\Box_\eta+{m_R}^2)^{-1}\mathcal{X}\,T\,.
\end{eqnarray}
Thus,  Eqs.~(\ref{PMfieldequationFOG32}) and (\ref{PMfieldequationFOG33}) have been decoupled. Let us now rewrite  Eq.~(\ref{PMfieldequationFOG43}) as:
\begin{eqnarray}\label{PMfieldequationFOG5}
(\Box_\eta+ m^2_+)(\Box_\eta+ m^2_{-})
 \varphi\,=\,-{m_R}^2f_{R\phi}(0,0,\phi^{(0)})\,\mathcal{X}\,T~;
\end{eqnarray}
\begin{equation}
\label{parameters}
m^2_{\pm}\equiv{m_R}^2\,w_{\pm}\bigl(\xi,\eta\bigl)~,
\end{equation}
where
\begin{eqnarray}\nonumber
w_{\pm}(\xi,\eta)&=&\frac{1-\xi+\eta^2\pm\sqrt{(1-\xi+\eta^2)^2-4\eta^2}}{2}\,,\\
\nonumber
\xi &=&3{f_{R\phi}(0,0,\phi^{(0)})}^2\,,\\
\nonumber 
\eta\,&=&\,\frac{m_\phi}{m_R}\,,
\end{eqnarray}

with the constraint $\xi<1$.

Applying on the first member of Eq.~(\ref{Position4}) the operators:
\begin{eqnarray}
\nonumber
(\Box_\eta+{m_Y}^2)\Box_\eta\,\,\,\,\,\,\text{and}\,\,\,\,\,\,
(\Box_\eta+ m^2_+)(\Box_\eta+ m^2_{-})\,, 
\end{eqnarray}
using Eqs. (\ref{PMfieldequationFOG41}), (\ref{PMfieldequationFOG42}) and (\ref{PMfieldequationFOG5}), we get
\begin{eqnarray}
\label{Position45}
(\Box_\eta+ m^2_+)(\Box_\eta+ m^2_{-})(\Box_\eta+{m_Y}^2)\Box_\eta h_{\mu\nu}= -2\,{m_Y}^2\,\mathcal{X}(\Box_\eta+ m^2_+)(\Box_\eta+ m^2_{-})\,T_{\mu\nu}
\nonumber\\
~~~~~~~~~~~~~~~~~~~~~~~~~+2\,m^2_R\,\mathcal{X}\,A(\partial)(\Box_\eta+ m^2_+)(\Box_\eta+ m^2_{-})T
\nonumber\\
~~~~~~~~~~~~~~~~~~~~~~~~~~~~~~+2\,{m_R}^4\,f_{R\phi}(0,0,\phi^{(0)})^2\,\mathcal{X}\,B(\partial)(\Box_\eta+{m_Y}^2)\Box_\eta T~.
\end{eqnarray}
Considering the associated homogeneous field equation:
\begin{eqnarray}
\label{HomogenFielEq}
(\Box_\eta+ m^2_+)(\Box_\eta+ m^2_{-})(\Box_\eta+{m_Y}^2)\Box_\eta h_{\mu\nu}=0\,,
\end{eqnarray}
we note that the dynamics of the perturbation $h_{\mu\nu}$ is made by four different modes: a massles one, which is the standard graviton, and three massive modes with masses $m_+$, $m_-$ and $m_Y$.  These quantities are free parameters which can be constrained via observational data. In our analysis, we suppose that all masses are different.

\section{Experimental constraints}\label{constraints}

For the graviton (massless mode) the dispersion relation is $E^2=p^2$, while for the three massive modes, $E^2=p^2+m^2_\alpha$, with $\alpha\equiv\{+,-,Y\}$, or equivalently:
\begin{eqnarray}
\label{DispersionRelation}
E^2=p^2+\,m_\alpha^2\,,\,\,\,\,\,\,\alpha\equiv\{+,-,Y\}\,.
\end{eqnarray}
The group velocity \cite{Will}, $v=\frac{p}{E}$ associated with the dispersion relation (\ref{DispersionRelation}) is:
\begin{eqnarray}
\label{groupvelocity}
v^g_\alpha=\sqrt{1-\frac{m_\alpha^2}{E^2}}\,,\,\,\,\,\,\,\alpha\equiv\{+,-,Y\}\,,
\end{eqnarray}
where $v$, $p$ and $E$ denote respectively, the velocity, momentum and energy. Note that if the rest mass vanishes, $m_\alpha=0$, the  group velocity is 1, and we have a massless particle, i.e. the graviton. Whereas, for $m_\alpha\neq 0$, the  group velocity is less than that of the graviton, since we have massive particle.

We will test the compatibility of Scalar Tensor Fourth Order Gravity  theories (\ref{FOGaction}) using current experimental data. We assume that the amplitude of massive modes is either comparable to that of the massless mode, or it lies below the detection threshold. Thus, experiments should be able to detect such massive modes independently of whether the massless mode has been detected.
Note that if the amplitude of the massive modes is smaller than that of the massless mode, one would not expect a detection and consequently there would be no constraints. 

Using the recent experimental results on the detection of gravitational waves (massless mode) by Advanced LIGO and Advanced Virgo \cite{CitGWTC-1,LIGOScientific:2019fpa} and
following the results presented in the reference \cite{LIGOScientific:2019fpa} we can impose the constraint on the three massive modes :
\begin{eqnarray}
\label{ConstrintCitGWTC-1}
m_\alpha<\,0.583\times 10^{-22}\,\text{eV}\,,\,\,\,\,\,\,\alpha\equiv\{+,-,Y\}\,.
\end{eqnarray}
For the next section, we indicate with $m_0$ the value of upper bound for the graviton mass, i.e. $m_0\equiv 0.583\times 10^{-22}\,\text{eV}=2.96\times 10^{-16}\,\text{m}^{-1}$.

\section{Scalar tensor fourth order gravity models}\label{STFOG_models}

\begin{center}
\begin{table*}[ht]
{\small
\hfill{}
\begin{tabular}{|l|l|c|c|c|c|c|c|}
\hline
\multicolumn{1}{|c|}{\textbf{Case}}&\multicolumn{1}{|c|}{\textbf{EGM}}& \multicolumn{6}{|c|}{\textbf{Parameters}}\\
\cline{3-8}
& & $m^2_R$ & $m^2_Y$ &$m^2_\phi$&$\xi$&$\eta$&$m^2_\pm$ 
 \\
\hline
\hline
A&\tiny{$f(R)$ }&\tiny{$-\frac{f_{R}(0)}{3f_{RR}(0)}$}&$\infty$& 0 & 0 & 0 & 0 
\\
\hline
\hline
B&\tiny{$f(R,R_{\alpha\beta}R^{\alpha\beta})$}&\tiny{$-\frac{f(0)}{3f_{RR}(0)+2f_Y(0)}$}&\tiny{$\frac{f_{R}(0)}{f_Y(0)}$}& 0 &0 & 0 & 0 
\\
\hline
\hline
C&\tiny{$f(R,\phi)+\omega(\phi)\phi_{;\alpha}\phi^{;\alpha}$}&\tiny{$-\frac{f_{R}(0)}{3f_{RR}(0)}$}& $\infty$ &\tiny{$-\frac{f_{\phi\phi}(0)}{2\omega(\phi^{(0)})}$}&\tiny{$\frac{3{f_{R\phi}(0)}^2}{2\omega(\phi^{(0)})}$}&\tiny{$\frac{m_\phi}{m_R}$}&\tiny{$m^2_R w_{\pm}(\xi,\eta)$}
\\
\hline
\hline
D&\tiny{$f(R,R_{\alpha\beta}R^{\alpha\beta},\phi)+\omega(\phi)\phi_{;\alpha}\phi^{;\alpha}$} &\tiny{$-\frac{f(0)}{3f_{RR}(0)+2f_Y(0)}$}&\tiny{$\frac{f_{R}(0)}{f_Y(0)}$}&\tiny{$-\frac{f_{\phi\phi}(0)}{2\omega(\phi^{(0)})}$}&\tiny{$\frac{3{f_{R\phi}(0)}^2}{2\omega(\phi^{(0)})}$}&\tiny{$\frac{m_\phi}{m_R}$}&\tiny{$m^2_R w_{\pm}(\xi,\eta)$} 
\\
\hline
\end{tabular}}
\hfill{}
\caption{\label{tab:tab2} Some models of extended theories of gravity.}
\label{tab:tab2}
\end{table*}
\end{center}

Let us consider some extended gravity models studied in the literature:

{\begin{itemize}

\item Case A: $f(R)$ denotes a family of theories, each one defined by a different function $f$ of the Ricci scalar $R$. The simplest case is for $f$ being equal to Ricci scalar, obtaining just General Relativity. In this case the characteristic scale (mass) $m_R$ (see, Table \ref{tab:tab2}, case A) depends only on the first and second derivatives of $f(R)$. Therefore, if we consider, for example, a polynomial expression $f(R)\,=\,R+\alpha\,R^2+\sum_{n\,=\,3}^{N}\alpha_n\,R^n$, only the terms $R$ and $R^2$ make contributions. Using the constraint Eq.~(\ref{ConstrintCitGWTC-1}), i.e.  $m_R<m_0$, we get the following relation between the derivates of $f(R)$:
 \[
-f_{RR}(0)>\frac{f_{R}(0)}{3\,m_0^2}.
\]
A particular case of  an $f(R)$ theory is the Starobinsky model, $f(R)\,=\,R-R^2/R_0$~\cite{Starobinsky}, with $R_0$ a constant. Here  the mass (\ref{mass_definition}) reads $m_R=\sqrt{R_0/6}$ ($m_Y=\infty,\, m_\phi=0$) and using Eq.~(\ref{ConstrintCitGWTC-1}) we obtain:
\begin{eqnarray}
 \label{r0}
R_0<6\,m_0^2\approx \, 5.3\times 10^{-31}\,\text{m}^{-2}.
\end{eqnarray}
However, at present the best constraint on $R_0$ is obtained by Solar System tests \cite{BerryGair} and reads $R_0>2.5\times 10^{8}\,\text{m}^{-2}$; a constraint with which (\ref{r0}) is not compatible. One may thus conclude that  the amplitude of massive modes is not detectable on Earth. Although the bounds we obtained are weaker, they could still be of interest, if for instance
there are modifications of the effective form of  $f(R)$ in different regions, through the chameleon mechanism.

\item Case B: $f(R,\,R_{\alpha\beta}R^{\alpha\beta})$, namely we also include the curvature invariant $R_{\alpha\beta}R^{\alpha\beta}$. In this case there are  two characteristic scales $m_R$ and $m_Y$. Using Eq.~(\ref{ConstrintCitGWTC-1}), $m_R<m_0$, and $m_Y<m_0$, with the mass definition Eqs.  (\ref{mass_definition}), we obtain the following relations between the derivatives:
 \begin{eqnarray}
 \label{dis}
 -f_{RR}(0,0)>\frac{2}{3}f_{Y}(0,0)+\frac{f_{R}(0,0)}{3\,m_0^2} \ \ , \ \
f_{Y}(0,0)>\frac{f_{R}(0,0)}{m_0^2}.
 \end{eqnarray}
As an illustration, let us consider $f(R)\,=\,R-R^2/R_0+R_{\alpha\beta}R^{\alpha\beta}/R_{ic}$, where $R_0$ and $R_{ic}$ are constants. The masses (\ref{mass_definition}) are then given by 
\begin{eqnarray}
\nonumber
m_R=\sqrt{\frac{R_0\,R_{ic}}{6R_{ic}-2R_0}}\,,\,\,\,\,
m_Y=\sqrt{R_{ic}}\,\,\,\,\text{and}\,\,\,m_\phi =0\,.
\end{eqnarray}
In this case the inequalities (\ref{dis}) become:
 \begin{eqnarray}
 \nonumber
  R_0<\frac{6m_0^2\,R_{ic}}{2m_0^2+R_{ic}}\ \ , \ \
 R_{ic}<m_0^2\,,
 \end{eqnarray}
implying
\begin{eqnarray}
\nonumber
R_0<2 \,m^2_0\approx\, 1.8\times 10^{-31}\,\text{m}^{-2}\,,\,\,\,\,\,\,
R_{ic}<m^2_0\approx 8.8\times 10^{-32}\,\text{m}^{-2}\,.
\end{eqnarray}
This class of theories includes the case of Weyl squared, \emph{i.e.} $C_{\mu\nu\rho\sigma}C^{\mu\nu\rho\sigma}\,=\,2R_{\mu\nu}R^{\mu\nu}-\frac{2}{3}R^2$, where there is only one characteristic scale since $m_R\,\rightarrow\,\infty$.\\

\item Case C: Scalar-tensor models $f(R,\,\phi)+\omega(\phi)\phi_{;\alpha}\phi^{;\alpha}$. There there are two effective scales $m_+$ and $m_-$ generated from the interaction between gravity and the scalar field (see, Table \ref{tab:tab2}, case C). Imposing Eq.~(\ref{ConstrintCitGWTC-1}),
$m_+<m_0$ and $m_-<m_0$, and after some algebra we get
 \begin{eqnarray}
 \label{C1}
m_R\,m_\phi <m_0^2\ \ &,& \ \
\bigg(1-\frac{m_0^2}{m_R^2}\bigg)\bigg(1- \frac{m_\phi ^2}{m_0^2}\bigg)< \xi \leqslant \bigg( \frac{m_\phi }{m_0}-1\bigg)^2\,,	\\
  \label{C2}
m_R\,m_\phi <m_0^2\ \ &,& \ \
0 <\xi \leqslant \bigg(\frac{m_\phi }{m_R}-1\bigg)^2 \,.
 \end{eqnarray}
Therefore, from Eqs.  (\ref{mass_definition}), (\ref{C1}) and (\ref{C2}) we find the following relations for the derivatives of the unspecified function  $f$:
 \begin{eqnarray}
\label{Rel12s}
\frac{f_{RR}(0,\,\phi^{(0)})}{f_{\phi\phi}(0,\phi^{(0)})} >\,\frac{f_{R}(0,\,\phi^{(0)})}{6\,\omega(\phi^{(0)})\, m_0^4}\,, 
\,\,\,\,\,\,\,\,
\Theta_C <3{f_{R\phi}^2(0,\,\phi^{(0)})}\leqslant\Lambda_C  \,,
 \end{eqnarray}
where
 \begin{eqnarray}
 \nonumber
\Theta_C &\equiv &  \bigg(1+\frac{3\,m_0^2\,f_{RR}(0,\,\phi^{(0)})}{f_{R}(0,\,\phi^{(0)})}\bigg)\bigg(1+\frac{f_{\phi\phi}(0,\phi^{(0)})}{2\omega(\phi^{(0)})\,m_0^2}\bigg)\,,\\\nonumber\\
\nonumber
\Lambda_C &\equiv &\bigg(\sqrt{\frac{-f_{\phi\phi}(0,\phi^{(0)})}{2\,\omega(\phi^{(0)})\,m_0^2}}-1\bigg)^2\,.
 \end{eqnarray}

As an example, let us consider the the following scalar-tensor theory:
\begin{eqnarray}\nonumber
f_{\rm ST}(R,\phi)\,=R-\frac{R^2}{R_0}+\lambda\,\varphi\,R+\frac{1}{2}\varphi_{,\alpha}\varphi^{,\alpha}-\frac{1}{2}{\cal M}^2\,\varphi^2~.
\end{eqnarray}
In this case, the masses (\ref{mass_definition}) are given by

\begin{eqnarray}
\nonumber
m_R=\sqrt{\frac{R_0}{6}}\,,\,\,\,\,
m_Y=\infty\,\,\,\,\text{and}\,\,\,m_\phi ={\cal M}\,.
\end{eqnarray}

The relations Eqs.~(\ref{Rel12s}) read
\begin{eqnarray}
  \nonumber
R_0 <\frac{6\, m_0^4}{{\cal M}^2}\,, \ \  
\bigg(1-\frac{6\,m^2_0}{R_0}\bigg)\bigg(1-\frac{{\cal M}^2}{m_0^2}\bigg)
< 3\,\lambda^2 \leqslant \bigg(\frac{{\cal M}}{ m_0}-1\bigg)^2\,.
 \end{eqnarray}
 
The first  inequality shows a simple relation between the scalar curvature $R_0$ and the mass of the scalar field ${\cal M}$.

\item Case D: $f(R,\,R_{\alpha\beta}R^{\alpha\beta},\phi)+\omega(\phi)\phi_{;\alpha}\phi^{;\alpha}$,  for which we have three effective scales $m_+$ and $m_-$ and $m_Y$. Imposing the constraints on all masses ($m_+<m_0$, $m_-<m_0$ and $m_Y<m_0$), we get the following relations
 \begin{eqnarray}
 \nonumber
m_R\,m_\phi <m_0^2\,&,& \ \
\bigg(1-\frac{m_0^2}{m_R^2}\bigg)\bigg(1- \frac{m_\phi ^2}{m_0^2}\bigg)< \xi \leqslant \bigg( \frac{m_\phi }{m_0}-1\bigg)^2\,,	
 \\
m_R\,m_\phi <m_0^2\, &,& \ \
  \nonumber
0 <\xi \leqslant \bigg(\frac{m_\phi }{m_R}-1\bigg)^2 \,,\\ \ \
  \nonumber
  m_Y< m_0&.&
 \end{eqnarray}

Using the definitions of the masses (\ref{mass_definition}), we get

 \begin{eqnarray}
\nonumber
&&\frac{3\,f_{RR}(0,0,\,\phi^{(0)})+2\,f_{Y}(0,0,\,\phi^{(0)})}{f_{\phi\phi}(0,0,\phi^{(0)})} >\,\frac{f_{R}(0,0,\,\phi^{(0)})}{2\,\omega(\phi^{(0)})\, m_0^4}\,,\\ \nonumber\\
\label{Rel145}
&&\Theta_D <3{f_{R\phi}^2(0,0,\,\phi^{(0)})}\leqslant\Lambda_D  \,,\\\nonumber\\
\nonumber
&&f_{Y}(0,0,\,\phi^{(0)})>\frac{f_{R}(0,0,\,\phi^{(0)})}{m_0^2}\,,
 \end{eqnarray}
where
 \begin{eqnarray}
 \nonumber
\Theta_D &\equiv & \bigg(1+\frac{m_0^2\,[3\,\,f_{RR}(0,0,\,\phi^{(0)})+2\,f_{Y}(0,0,\,\phi^{(0)})]}{f_{R}(0,0,\,\phi^{(0)})}\bigg)\bigg(1+\frac{f_{\phi\phi}(0,0,\phi^{(0)})}{2\omega(\phi^{(0)})\,m_0^2}\bigg)\,,\\\nonumber\\
\nonumber
\Lambda_D &\equiv &\bigg(\sqrt{\frac{-f_{\phi\phi}(0,0,\phi^{(0)})}{2\,\omega(\phi^{(0)})}}-1\bigg)^2\,.
 \end{eqnarray}

In this case as example we consider the following model:
\begin{eqnarray}
  \nonumber
f(R,R_{\alpha\beta}R^{\alpha\beta},\phi)\,=R-\frac{R^2}{R_0}+\frac{R_{\alpha\beta}R^{\alpha\beta}}{R_{ic}}+\lambda\,\varphi\,R+\frac{1}{2}\varphi_{,\alpha}\varphi^{,\alpha}-\frac{1}{2}{\cal M}^2\,\varphi^2~,
\end{eqnarray}
with masses
 \begin{eqnarray}
\nonumber
m_R=\sqrt{\frac{R_0\,R_{ic}}{6R_{ic}-2R_0}}\,,\,\,\,\,
m_Y=\sqrt{R_{ic}}\,\,\,\,\text{and}\,\,\,m_\phi ={\cal M}\,.
\end{eqnarray}

The relations (\ref{Rel145})  read
\begin{eqnarray}
  \nonumber
&&R_0\, <\frac{\,6\, m_0^4\,R_{ic}}{2\,m_0^4+R_{ic}\,{\cal M}^2}\,,\\ \nonumber\\ 
  \nonumber
&&\bigg(1-2\frac{(3R_{ic}-R_0)\,m^2_0}{R_0\,R_{ic}}\bigg)\bigg(1-\frac{{\cal M}^2}{m_0^2}\bigg)
< 3\,\lambda^2 \leqslant \bigg(\frac{{\cal M}}{ m_0}-1\bigg)^2\,,\\\nonumber\\
\nonumber
&&R_{ic}<m_0^2\,,
 \end{eqnarray}
from which we derive the following inequality between the scalar curvature $R_0$ and the mass of the scalar field ${\cal M}$:  
 
 \begin{eqnarray}
  \nonumber
R_0\, <\frac{\,6\, m_0^4}{2\,m_0^2+{\cal M}^2}\,.
 \end{eqnarray}

In general, the constraints one can obtain are rather weak, see Eq.~(\ref{r0}), but can in principle be improved once further data of nearby  binary black hole signals are available.

\end{itemize}}

\section{Conclusions}\label{conclusions}

We have  studied the propagation of gravitational waves in the context of fourth order gravity nonminimally coupled to a massive scalar field, in the weak-field approximation. 

We have seen that, in general, extended models of gravity have some massive modes in addition to the massless mode (graviton) of General Relativity. In our analysis, we assumed that these modes has the amplitude comparable to that of the massless mode. Using the recent observations of gravitational radiation from the binary black hole signals detected by Advanced LIGO and Advanced Virgo, published in the catalog GWTC-1 \cite{LIGOScientific:2018mvr}, we have constrained the free parameters of extended gravity models.
These constraints  are weaker compared to the ones obtained from experimental data of a different origin \cite{gravito_mag}. 

For the $f(R)$-theories we have analysed, the constraint derived from gravitational waves is not compatible with the one currently obtained from Solar System experiments. Therefore,  the amplitude of massive modes cannot be detected on Earth. This could happen because, these massive modes are either significantly suppressed by a small coupling parameter or by a large energy scale that affects them.

Nevertheless, the analysis presented here is important for two reasons. Firstly, there could exist a, yet unknown, screening mechanism operating on Earth and Solar System scales, (for instance the chameleon screening \cite{veltman,symmetron,sakstein,brax}), but could not manifest on larger (astrophysical) scales. In fact, one may expect that the variation of speed of GWs might occur in high density environments, but such a screening effect should be reduced over distances of order 40 Mpc \cite{creminelli,piazza1}.
Secondly, the approach we have followed here can be used once further data of nearby binary black hole signals are available, probably leading to stronger constraints.  
Moreover, further observations over larger distances could provide limits on both screening mechanisms and higher derivative corrections, in particular on the effective model used in the analysis presented here.

\acknowledgments
M.S. is supported in part by the Science and Technology Facility Council (STFC), United Kingdom, under the research grant ST/P000258/1. The authors would like to thank the anonymous referee
for observations that improved the manuscript.

\end{document}